\documentclass[letterpaper, 10 pt, conference,fleqn]{ieeeconf}  % Comment this line out if you need a4paper

\IEEEoverridecommandlockouts                              

\usepackage{amsmath,amssymb,amsfonts}
\usepackage{graphicx}
\usepackage{xcolor}                   % (optional) but before hyperref
\usepackage[caption=false,font=footnotesize]{subfig}
\let\labelindent\relax
\usepackage{enumitem}
\usepackage{algorithm}
\usepackage{algorithmic}

\usepackage{amsthm}
\allowdisplaybreaks
\usepackage{textcomp}
\usepackage{cite}                     
\usepackage{hyperref}                 % LAST
\hypersetup{hidelinks=true}
\usepackage{mathtools}
\newtheorem{definition}{Definition}
\newtheorem{lemma}{Lemma}
\newtheorem{theorem}{Theorem}

\usepackage{cleveref}
\usepackage{etoolbox}
\newtheorem*{assumpblock}{Assumptions}

\newlist{assumplist}{enumerate}{1}
\setlist[assumplist]{label=(A\arabic*), ref=A\arabic*, leftmargin=*, itemsep=0.25em}

\crefname{assumplisti}{Assumption}{Assumptions}

\title{\LARGE \bf
	Steering Opinion Dynamics in Signed Time-Varying Networks via External Control Input*
}

\author{Swati Priya$^{1}$ and Twinkle Tripathy$^{2}$% <-this % stops a space
	\thanks{*This work was not supported by any organization}% <-this % stops a space
	\thanks{$^{1}$Swati Priya is a graduate student in the Department of Electrical Engineering, Indian Institute of Technology, Kanpur, Uttar Pradesh, India
		{\tt\small swatipr21@iitk.ac.in}}%
	\thanks{$^{2}$Twinkle Tripathy is Assistant Professor with the Control and Automation specialization of the Department of Electrical Engineering, Indian Institute of Technology Kanpur, Uttar Pradesh, India
		{\tt\small ttripathy@iitk.ac.in}}%
}

\begin{document}

	\maketitle
	\thispagestyle{empty}
	\pagestyle{empty}

	%%%%%%%%%%%%%%%%%%%%%%%%%%%%%%%%%%%%%%%%%%%%%%%%%%%%%%%%%%%%%%%%%%%%%%%%%%%%%%%%
	
	\begin{abstract}
		This paper studies targeted opinion formation in multi-agent systems evolving over signed, time-varying directed graphs. The dynamics of each agent's state follow a Laplacian-based update rule driven by both cooperative and antagonistic interactions in the presence of exogenous factors. We formulate these exogenous factors as external control inputs and establish a suitable controller design methodology enabling collective opinion to converge to any desired steady-state configuration, superseding the natural emergent clustering or polarization behavior imposed by persistently structurally balanced influential root nodes. 
        Our approach leverages upper Dini derivative analysis and Grönwall-type inequalities to establish exponential convergence for opinion magnitude towards the desired steady state configuration on networks with uniform quasi-strong $\delta$-connectivity. Finally, the theoretical results are validated through extensive numerical simulations. 
	\end{abstract}
%%%%%%%%%%%%%%%%%%%%%%%%%%%%%%%%%%%%%%%%%%%%%%%%%%%%%%%%%%%%%%%%%%%%%%%%%%%%%%%%
	\section{INTRODUCTION}

The ubiquity of digital connectivity has transformed social networks into powerful platforms for shaping and disseminating opinions. This has spurred a surge of interest in opinion dynamics, where researchers investigate phenomena like opinion clustering and polarization using agent-based models. Studies suggest that beyond interpersonal influence, opinion formation today is increasingly driven by external forces such as mass media effects
\cite{sirbu2016opinion} and targeted media campaigns \cite{kamthe2025,priya,covid}. Empirical evidence indicates that both the traditional media \cite{baum2008new} and social media
\cite{denniss2025social,tucker2018social} are frequently exploited to manipulate public beliefs by flooding users with misinformation and partisan content. This leads to individual opinions moving toward more extreme viewpoints. For instance, during the COVID-19 pandemic, social media platforms were inundated with false claims about vaccines, which polarized communities into anti- and pro-vaccine camps and undermined public health efforts by government, contributing to widespread vaccine hesitancy \cite{denniss2025social}. With the rapid rise of generative AI technologies (e.g., ChatGPT and other chatbots) that automatically generate persuasive content, opinion manipulation has become more potent than ever. These factors motivate the study of targeted opinion formation in both online and offline social networks, aiming to steer public discourse toward evidence-based consensus while countering the harmful effects of polarization.

Some recent works \cite{covid,kamthe2025,priya} introduce external control inputs into agent-based opinion models, which exploit knowledge of the network structure to steer agents’ states toward favorable outcomes, particularly in public-health applications such as vaccine promotion and misinformation control. In \cite{covid}, media campaign effects are incorporated into Friedkin–Johnsen dynamics, and targeted pro-vaccine campaigns are designed to increase the expected fraction of individuals willing to be vaccinated. In \cite{kamthe2025}, a Laplacian-based opinion model with cooperative interactions and an external bias term is considered, and a suitably designed control input is shown to steer agent opinions to any desired steady state. Furthermore, in our earlier work \cite{priya}, we proposed a controller design for arbitrary signed networks that steers agents’ opinions to any desired targeted state. However, these results focus on static network topologies, whereas real social networks exhibit continuously evolving trust (cooperative) and mistrust (antagonistic) relationships. This motivates the study of targeted opinion formation on signed graphs with time-varying interaction topology.

 Several pioneering works \cite{altafini,proskurnikov2015,xia2015structural} study opinion polarization and clustering over time-varying directed signed graphs. These studies identify connectivity and signed interaction patterns under which agents’ opinions split into distinct clusters. For instance, \cite{altafini} examines modulus consensus (opinions equal in magnitude but opposite in sign) on time-varying directed signed graphs and shows that uniform strong connectivity together with structural balance leads to bipartite consensus (opinion polarization). The work \cite{proskurnikov2015} further establishes necessary and sufficient conditions for bipartite consensus (resp. stability) over cut-balanced graphs with switching topology, requiring essential strong connectivity and structural balance (resp. unbalance). Later, \cite{xia2015structural} showed that, in time-varying signed networks, DeGroot-type averaging drives opinions toward polarization or clustering if, over every infinite sequence of uniformly bounded nonempty intervals, the union graph contains a directed spanning tree and the root node subgraph is structurally balanced. However, these clustering outcomes are entirely driven by endogenous agent interactions and do not account for exogenous control inputs that could steer the collective state away from its natural emergent behavior.

For targeted opinion formation on time-varying networks, the literature is still quite limited. For example, the paper authors in \cite{zou2020targeted} formulate a constraint-based targeted bipartite consensus problem in signed networks with switching topologies, where each agent is given a credibility interval (convergence set) for its opinion. Under structural‐balance and uniform joint connectivity assumptions on the time-varying graph, they prove that agents reach the desired bipartite consensus, with each agent's opinion converging within its credibility interval. Using a game-theoretic framework, \cite{lang2023time} study competitive targeted marketing on social networks, where firms allocate limited marketing resources as exogenous inputs to network agents (customers) in order to sway their opinions and maximize finite-horizon profit. Other recent works achieve target consensus by altering the network structure rather than directly nudging opinions. The paper \cite{musco2018minimizing} minimizes polarization via graph edge rewiring, while \cite{steering} proposes a coevolutionary model where both opinions and edge weights evolve under a dynamic control input to reach the desired consensus state. 

Motivated by these works, we propose a decentralized controller design approach that utilizes the knowledge of the network structure for effective targeted opinion shaping in social networks with signed time-varying interaction topology. Building on our prior controller design for static signed graphs \cite{priya}, we now consider switching topologies and design control inputs that steer the agents’ opinions to prescribed target states. Such structure-aware control opens new possibilities for real-world interventions, for example, guiding public-health awareness campaigns or countering online misinformation, by systematically steering collective opinions beyond the network’s natural emergent behavior.

The remainder of this paper is structured as follows. Section~\ref{pre} introduces the necessary notations and preliminaries. Section~\ref{problem} formulates the problem under consideration. Section~\ref{lemma} presents the controller design methodology, including key lemmas and the main theorem. Numerical simulations illustrating the effectiveness of the proposed approach are given in Section~\ref{sim}. Section~\ref{disc} discusses potential real-world implications of the proposed controller design methodology, and Section~\ref{con} concludes the paper.

	\section{Notations and Preliminaries}\label{pre}
	\subsection*{Notations}
Let \( \mathbb{R}^{n \times m} \) denote the space of real \( n \times m \) matrices, \( \mathbb{R} \) the set of real numbers, and \( \mathbb{Z}_{\geq 0} \) the set of non-negative integers. The sign function $\operatorname{sgn}(x) \in \{-1, 0, 1\} $ denotes the sign of a number x. The notation \( |\cdot| \) denotes absolute value. For a set \( S \), its cardinality is given as \( \operatorname{card}(S) \). The abbreviations “a.a.” and “a.e.” stand for “almost all” and “almost everywhere,” respectively. Let \( \mathcal{N}_i \) denote the set of incoming neighbors of agent \( i \) ( i.e., the set of agents that have a directed edge toward \( i \).

\subsection*{Preliminaries}

A signed digraph \( \mathcal{G} = (V, E, A) \) consists of a node set \( V = \{1, \dots, N\} \), edge set \( E \subseteq V \times V \), and a weighted adjacency matrix \( A \in \mathbb{R}^{n \times n} \), where \( a_{ij} \ne 0 \iff (v_j, v_i) \in E \). The edge \( (v_j, v_i) \in E \) indicates information flow from agent \( j \) to agent \( i \).

A subgraph \( \mathcal{G}' = (V', E') \) of the graph $\mathcal{G}$ satisfies  \( V' \subseteq V \), \( E' \subseteq E \cap (V' \times V') \). A directed path of length \( n-1 \) is a sequence of consecutive directed edges \( (v_{i_1}, v_{i_2}), \dots, (v_{i_{n-1}}, v_{i_n}) \in E \). If all the nodes are distinct, it is termed a \emph{simple path}. Define \( d_{\max}(i, j) \) as the length of the longest simple path from \( i \) to \( j \); then the longest path length of the graph $\mathcal{G}$ is defined as $D_{\max}(\mathcal{G}) := \max_{i, j \in V} d_{\max}(i, j).$

A node is called a \emph{root} node if it can reach all other nodes via directed paths. A digraph is \emph{strongly connected} if every node is a root node. A digraph is said to be quasi-strongly connected if it contains at least one root node. A \emph{directed tree} is the minimal quasi-strongly connected graph where there is exactly one root, and every other node can be reached from it through a unique directed path. A \emph{directed spanning tree} in a graph is a directed tree contained in the graph that includes all the nodes of the graph.

A \emph{strongly connected component} (SCC) is a maximal subgraph that is strongly connected and is not
contained in any larger strongly connected subgraph of a graph. The \emph{condensation graph} of \( \mathcal{G} \), denoted by \( \mathcal{C(G)} \), is a directed graph in which each node represents an SCC of \( \mathcal{G} \), with edges indicating interaction between two SCCs. An SCC with no incoming edges from other SCCs is called a \emph{closed} SCC.
    
A static signed digraph is \emph{structurally balanced} if the vertex set \( V \) can be partitioned into two disjoint subsets \( V_1 \) and \( V_2 \) such that all edges within each subset are positive and all edges between the subsets are negative. Otherwise, the digraph is \emph{structurally unbalanced}.

The \emph{upper Dini derivative} \cite{lin2007state,danskin1966theory} of a real-valued function \( h(t) \) is defined as: $D^+ h(t) = \limsup_{s \to 0^+} \frac{h(t + s) - h(t)}{s}.$

	\section{Problem statement} \label{problem}
	\subsection{Network dynamics and  assumptions}
	We consider a multi-agent system composed of $N$ agents indexed by the set \(V = \{1, \dots, N\}\), with \(N \ge 2\). Each agent \(i \in V\) represents a state \(x_i(t) \in \mathbb{R}\) at time \(t\). The system's evolution begins at an initial time \(t_0 \ge 0\), and the state of each agent evolves according to the following dynamics:
    \begingroup
\setlength{\abovedisplayskip}{2pt}
\setlength{\belowdisplayskip}{2pt}
	\begin{equation}
		\dot{x}_i(t) = \sum_{j \in \mathcal{N}_i} |a_{ij}(t)| \left( \text{sgn}(a_{ij}(t)) x_j(t) - x_i(t) \right) + u_i,
		\label{system}
	\end{equation}
    \endgroup
	where \( a_{ij}(t) \in \mathbb{R} \) denotes the strength of the interaction from agent \( j \) to agent \( i \), and \( \mathcal{N}_i \) is the set of neighbors of agent \( i \). 
	
	In compact matrix form, the dynamics can be written as:
	\begin{equation}
		\dot{\mathbf{x}}(t) = -L(t)\mathbf{x}(t) + \mathbf{u}(t),
		\label{system_1}
	\end{equation}
	where \( L(t) \in \mathbb{R}^{N \times N} \) is the time-varying signed Laplacian matrix with elements defined by:
	\[
	[L(t)]_{ij} = 
	\begin{cases}
		\sum\limits_{k \in \mathcal{N}_i} |a_{ik}(t)|, & \text{if } i = j, \\[1ex]
		- a_{ij}(t), & \text{if } i \ne j.
	\end{cases}
	\]

\begin{assumpblock}\leavevmode
	\begin{assumplist} \label{assump}
		\item For all \(i,j\in V\), \(a_{ij}(t)\) is piecewise continuous on every compact interval with the set of discontinuities having Lebesgue measure zero.
		\item \(a_{ii}(t)\equiv 0\) for all \(i\in V\).
		\item There exists \(M_0>0\) s.t. \(\int_{t_1}^{t_2} |a_{ij}(s)|\,ds \le M_0 (t_2-t_1)\).
	\end{assumplist}
\end{assumpblock}
    	
	\subsection{Arcs, graph, and connectivity}
	The node dynamics in eqn. \eqref{system} gives rise to a time-varying directed communication graph, defined below.
	
	% --- Underlying communication graph
	\begin{definition}
		\label{def:comm_graph}
		Given system dynamics \eqref{system}, the underlying communication graph at time \(t\) is defined as a weighted digraph \(\mathcal G_A(t)=(V,E_t,A(t))\), where \((i,j)\in E_t\) iff \(a_{ij}(t)\neq 0\),
		and the edge weight is \(A_{ij}(t)=a_{ij}(t)\).
	\end{definition}
	
	\begin{definition}
		\label{def:delta_arc}
		Let  \(\delta>0\) and \(T>0\) be a given constant. 
		\begin{enumerate}[label=(\alph*)]
			\item \(\delta\)-arc- An arc \((i, j)\) is called a \(\delta\)-arc of the time-varying graph \( \mathcal{G}_A(t) \) over the interval \([t_1, t_2)\) if 
             \begingroup
            \setlength{\abovedisplayskip}{2pt}
            \setlength{\belowdisplayskip}{2pt}
			\[
			\int_{t_1}^{t_2} |a_{ij}(t)|\, dt \ge \delta (t_2 - t_1).
			\]
			\endgroup
			\item \(\delta\)-path- A path is a \emph{\(\delta\)-path} on \([t_1,t_2)\) if every arc in the path is a \(\delta\)-arc on \([t_1,t_2)\).
			
			\item Union and condensation graph- The graph $G_\delta[t_1,t_2))$ represents the union of all the $\delta$-arcs that appear over the interval $[t_1,t_2)$. The graph $G_\delta^\infty$ denotes the union of all the $\delta$-arcs that appear over the time interval $[0,\infty)$. Then, $\mathcal{C(G_\delta}[t_1,t_2)))$ and $\mathcal{C(G_\delta}[0,\infty)))$  represents the condensation graph associated with the union graph $G_\delta[t_1,t_2))$ and $G_\delta^\infty$, respectively. 
		\end{enumerate}
	\end{definition}
	
	\begin{definition}
		\label{def_uniformly}
		A time-varying digraph \(\mathcal{G}_A(t)\) is said to be \emph{uniformly quasi strongly \(\delta\)-connected} if there exists \(T>0\) such that, for every \(t \geq 0\), the condensation digraph \(\mathcal{C}(G_\delta[t,t+T))\) of the union graph \(G_\delta[t,t+T)\) contains at least one root node such that there exists a \(\delta\)-path from that node to every other node in \(\mathcal{C}(G_\delta[t,t+T))\).
	\end{definition}

	\begin{definition}
		\label{def:psb}
		Let the interaction digraph $G_{A(t)}$ be uniformly quasi strongly $\delta$-connected such that the condensation graph $\mathcal{C}(G^{\infty}_{\delta})$ contains a fixed set of nodes represented as $S$, that forms a closed strongly connected component (SCC) of $\mathcal{C}(G^{\infty}_{\delta})$. Then, the closed SCC is said to be \textbf{persistently structurally balanced} if for all $t\ge 0$, there exists a unique bipartition $\mathcal{S}=V^{s}_{1}\cup V^{s}_{2}$ with $V^{s}_{1}\cap V^{s}_{2}=\varnothing$
		such that $a_{jk}(t)\ge 0,$ if $j,k\in V^{s}_{1}$ or $j,k\in V^{s}_{2}$, and $a_{jk}(t)\le 0,$ if $j\in V^{s}_{1},\,k\in V^{s}_{2}$ or $j\in V^{s}_{2},\,k\in V^{s}_{1}$.
		
		The closed SCC $\mathcal{S}$ is \textbf{structurally unbalanced} if no such umique bipartition $(V^{s}_{1},V^{s}_{2})$ exists for all $t\ge 0$.
	\end{definition}
	\subsection{Targeted Opinion Formation over Time-Varying Graphs}
	
	We consider the network dynamics given in eqn. \eqref{system} over a time–varying interaction
	graph $\mathcal G_{A(t)}$ that is uniformly quasi strongly $\delta$–connected such that in every time interval $[t,t+T]$, for any $t \geq0$, the condensation graph $\mathcal{C(G_\delta}[t,t+T)))$ contains a fixed set of nodes $S\subseteq V$ that forms a closed SCC. The nodes in $S$ constitute an influential node (root node) set, from which there exists a $\delta$–path to every other node in the graph $\mathcal{G}_{\delta}[t,t+T)$. Several works \cite{xia2015structural,proskurnikov2015} show that the presence of such influential nodes, together with the structural balance of the subgraph formed by those nodes, can lead to opinion polarization or clustering among the agents. Our goal is to design a control input \( \mathbf{u}(t) \) such that the agent states asymptotically converge to a prescribed target state \( \mathbf{x}_d \in \mathbb{R}^N \), in the presence of such influential node i.e.,
     \begingroup
\setlength{\abovedisplayskip}{2pt}
\setlength{\belowdisplayskip}{2pt}
	\begin{equation*}
	\lim_{t \to \infty} \mathbf{x}(t) = \mathbf{x}_d.
	\end{equation*}	
    \endgroup

	\section{Controller design methodology}\label{lemma}
In this section, we develop a controller design methodology that steers the agents' states toward an arbitrary desired target over a time-varying interaction graph. The main result is stated in the form of a theorem. Before presenting this theorem, we first introduce several key lemmas that are instrumental in its proof.
	\begin{lemma}\label{lem:abs-dynamics}
		Let the time-varying interaction graph \(\mathcal{G}_A(t)\) be uniformly quasi strongly \(\delta\)-connected such that assumptions (A1)-(A3) holds.
		Then for the system \eqref{system_1} in the absence of control input i.e., $\mathbf{u}(t)\equiv 0$, the following dynamics holds $\forall i\in V$ and  for a.a. (almost all) \(t\geq 0\) 
       \begin{equation}\label{eq:abs-dyn}
			\frac{d}{dt}\,|x_i(t)| \;=\; \sum_{j\in\mathcal{N}_i} |a_{ij}(t)|\Bigl(\theta_{ij}(t)\,|x_j(t)| - |x_i(t)|\Bigr),
		\end{equation}
		where $\theta_{ij}(t):=\operatorname{sgn}(x_i(t))\operatorname{sgn}(a_{ij}(t))\operatorname{sgn}(x_j(t))\in\{-1,0,1\}$.    
	\end{lemma}
	\begin{proof}
   Under assumptions (A1)–(A3), the system \eqref{system_1} with $\mathbf{u}\equiv 0$, i.e., $\mathbf{\dot{x}} = -L(t)\mathbf{x}$ has time varying coefficients $a_{ij}(\cdot)$ that are measurable in $t$, integrably bounded and for each fixed $t$, $\mathbf{\dot{x}}$ is linear in $\mathbf{x}$.  So the vector field $X(t,x)=-L(t)\mathbf{x}$ satisfies the Carathéodory conditions \cite{filippov2013,discontinuous} and for every initial state there exists a unique absolutely continuous solution $\mathbf{x}(t)$ on $[0,\infty)$. In particular, each $x_i(t)$ is absolutely continuous (hence differentiable almost everywhere).

    Since $L(t)$ is piecewise-continuous, $\mathbf x(t)$ is actually Lipschitz (hence absolutely continuous) on any finite interval. Because $x_i(t)$ is absolutely continuous, and the scalar function $f(x)=|x|$ is Lipschitz (with constant 1), the composition $|x_i(t)|=f(x_i(t))$ is also absolutely continuous \cite{filippov2013}. It follows that $|x_i(t)|$ has a derivative for almost every $t \geq 0$. Moreover, whenever $x_i(t)\neq 0$, the chain rule of differentiation yields $\frac{d}{dt}|x_i(t)| = f'(x_i(t))\,\dot{x}_i(t) = \operatorname{sgn}(x_i(t))\,\dot{x}_i(t)$.
Substituting  $x_i$ from eqn. \eqref{system} gives  exactly the formula in~\eqref{eq:abs-dyn}.

Finally, if $x_i(t)=0$ on a set of positive measure (e.g. on an interval), then $\dot x_i(t)=0$ and we have $\frac{d}{dt}|x_i(t)|=0$. One may consistently set $\theta_{ij}(t)=0$ on that set. In this case both sides of \eqref{eq:abs-dyn} vanish a.e., so the equality still holds. 
    \end{proof}

	\begin{lemma}\label{max_error_dynamics}
Consider the system \eqref{system_1} with $\mathbf{u}(t) \equiv 0$, evolving over a signed time-varying graph $\mathcal{G}_A(t)$ that satisfies assumptions (A1)–(A3). Suppose that $\mathcal{G}_A(t)$ is uniformly quasi-strongly $\delta$-connected, i.e., there exists a constant $T>0$ such that, for every $t \ge t_0 \ge 0$, the graph $\mathcal{G}_\delta[t,t+T)$ has a condensation graph $\mathcal{C}(\mathcal{G}_\delta[t,t+T))$ containing a fixed node set $S \subseteq V$ forming an SCC. The nodes in $S$ also constitute a root node set, in the sense that there exists a $\delta$-path from $S$ to every other node in the receiver set $R := V \setminus S$ in the graph $\mathcal{G}_\delta[t,t+T)$ . Define the following:
		\begin{align}
			h(t) &:= \max_{i \in R} |x_i(t)|, \label{h}\\
			c(t) &:= \max_{i \in S} |x_i(t)|,  \label{c}
		\end{align}
		Then, for a.a. $t \geq t_0\geq 0$, the following inequalities hold:
         \vspace{-0.1 cm}
		\begin{align}
			D^+ h(t) &\leq M_0 N_s (c(t) - h(t)), \label{Dini_h} \\
			D^+ c(t) &\leq 0,  \label{Dini_c}
		\end{align}
		 where $D^+$ denotes the upper Dini derivative and $M_0 > 0$ is a constant defined earlier.
	\end{lemma}
    \vspace{ -0.4cm}
	\begin{proof}
		We first establish inequality \eqref{Dini_h}. Since each $|x_i(t)|$ is absolutely continuous, its derivative exists a.e. (except on a set of measure zero). Define the  index set $I(t) := \{ i \in R \;|\; |x_i(t)| = h(t) \}.$ 
		By the standard results on upper Dini derivatives (Lemma~2.1 in~\cite{shi2013robust}), we can write
		\begin{equation*}
		D^+ h(t) = \max_{i \in I(t)} \frac{d}{dt}|x_i(t)| \quad \text{for a.a. } t \geq t_0 \geq 0.
		\end{equation*}
		
Note that the neighbor set for an agent $i$  splits disjointly as  $\mathcal N_i=(\mathcal N_i\cap S)\cup(\mathcal N_i\cap R)$.
For $i\in I(t)$ we have $|x_i(t)|=h(t)$, and
$\theta_{ij}(t)|x_j(t)|\le c(t)$ if $j\in S$, while
$\theta_{ij}(t)|x_j(t)|\le h(t)$ if $j\in R$. Hence, from eqn. \eqref{eq:abs-dyn}, we get 
		\begin{align*}
			D^+ h(t) 
			&= \max_{i \in I(t)} \frac{d}{dt}|x_i(t)| \\
			&\leq \max_{i \in I(t)} \Bigg[ \sum_{j \in \mathcal N_i\cap S} |a_{ij}(t)| c(t) + \sum_{j \in \mathcal N_i\cap R} |a_{ij}(t)| h(t) \\
			& - \sum_{j \in \mathcal {N}_i}|a_{ij}(t)| h(t) \Bigg] \\
			&\leq \max_{i \in I(t)} \Bigg[ \sum_{j \in \mathcal {N}_i \cap S} |a_{ij}(t)| \big(c(t) - h(t)\big)  \Bigg].
		\end{align*}
    Since $|a_{ij}|$ is locally integrable by assumption (A3), the Lebesgue–Besicovitch Differentiation Theorem \cite[Thm.~1, \S 1.32]{evans2018measure}) gives the pointwise bound as $|a_{ij}(t)|\le M_0$ for a.a.\ $t\geq t_0$. Summing over all $j\in (\mathcal N_i\cap S)$ yields the following upper bound 
\begin{equation*}\label{eq:sumS-pointwise}
			\sum_{j\in \mathcal N_i\cap S} |a_{ij}(t)|  \leq	\sum_{j\in S} |a_{ij}(t)| \le M_0 N_S \qquad \text{for a.a.\ } t\ge t_0,
		\end{equation*}
		where $N_S := \operatorname{card}(S)$. This proves the inequality given in eqn. \eqref{Dini_h}.\\
	We now prove the inequality \eqref{Dini_c}.  
		For the given max function as defined in eqn.  \eqref{c}, let us define the index set $I_c(t) := \{ i \in S \;|\; |x_i(t)| = c(t) \}.$
		By the same argument as stated earlier, we can write
		\begin{align*}
			D^{+} c(t) &= \max_{i \in I_c(t)} \frac{d}{dt} |x_i(t)|, \quad \text{for a.a. } t \geq t_0 \geq 0.\\
			&=\max_{i \in I(t)} \Bigg[ \sum_{j \in \mathcal{N}_{i}} |a_{ij}(t)| \big(\theta_{ij}(t)|x_j(t)| - c(t)\big) \Bigg].
		\end{align*}
		Since the set $S$ forms a  closed SCC in the union graph $G_\delta[t,t+T)$, over  every time interval $[t,t+T)$ for any $t \geq 0$, hence  any  neighbor of an agent $i\in S$, always lies in the same set.  Thus  for any
		$j\in\mathcal N_i$ where $i \in S$, we get $ \theta_{ij}(t)\,|x_j(t)| \leq c(t)$. Consequently,
		$D^{+} c(t) \leq 0$. This completes the proof.
	\end{proof}
	\vspace{-0.5 cm}
	\begin{lemma}\label{k_times}
		Suppose the statement of  Lemma \ref{max_error_dynamics} holds and the
		inequalities \eqref{Dini_h} and \eqref{Dini_c} are satisfied. Define $s,k\in\mathbb{Z}_{\ge 0}$, and $K:=kT$. Then, for any
		$t_0\ge 0$ and for a.a.\ $t\in [\,t_0+sK,\; t_0+(s+1)K\,]$, the following bounds hold:
		\begin{align}
			c(t) &\le c(t_0+sK), \label{eq_c_bound}\\[1ex]
			h(t) &\le h(t_0+sK) + c(t_0+sK). \label{eq_h_bound}
		\end{align}
	\end{lemma}	
	\begin{proof}
		We first prove the inequality given eqn. \eqref{eq_c_bound}. Recall that from eqn. \eqref{Dini_c} we get $D^+ c(t) \le 0$ for a.a. $t \geq t_0 \geq 0$. Thus, $c(t)$ is non-increasing in $t$, and  hence for a.a. $t \in [t_0+sK,t_0+(s+1)K]$ we clearly obtain  $c(t) \leq c(t_0+sK)$.  
		
		We now discuss the proof for the inequality given in eqn. \eqref{eq_h_bound}. Recall from  eqn. \eqref{Dini_h} that for a.a $t\geq t_0 \geq 0$, we get
		\[
		D^+ h(t) \leq M_s\big(c(t) - h(t)\big), 
		\qquad \text{where }M_s := M_0 N_s.
		\]
By applying Grönwall’s inequality and using the bound $c(t) \leq c(t_0+sK)$ as mentioned in eqn.~\eqref{eq_c_bound}, it turns out that for a.a.  $t \in [t_0+sK,t_0+(s+1)K]$, 
		\begin{align}
			h(t) &\leq e^{-M_s(t-t_0-sK)} h(t_0+sK) +\nonumber\\
			& \left(1 - e^{-M_s(t-t_0-sK)}\right) c(t_0+sK).
		\end{align}
		The fact that $0 \leq e^{-M_s(t-t_0-sK)} \leq 1$, implies that
		$h(t) \leq h(t_0+sK) + c(t_0+sK)$, which proves eqn. \eqref{eq_h_bound}.
	\end{proof}
We now state the main theorem, which specifies a suitable control law for the time-varying interaction graph $\mathcal{G}_{A(t)}$.
\begin{theorem}\label{thm_1}
Let $\mathcal{G}_{A(t)}$ be a time-varying signed interaction graph that is uniformly quasi-strongly $\delta$–connected and satisfies assumptions (A1)–(A3). Suppose that for any $t\geq 0$, the union graph $\mathcal{G}_\delta[t,t+T)$ contains a fixed node set $S \subseteq V$ that forms a root node set, and that the subgraph induced by $S$ is persistently structurally balanced. Then, the system \eqref{system_1} with control input 
\begin{equation}
\mathbf{u}(t) = L(t)\mathbf{x}_d - K(t)\bigl(\mathbf{x}(t) - \mathbf{x}_d\bigr), 
\label{u}
\end{equation} 
with $K(t) = \mathrm{diag}\bigl(k_1(t),\dots,k_N(t)\bigr),$
achieves the desired steady state $\mathbf{x}_d$, provided that:
\begin{enumerate}[label=(P\arabic*), leftmargin=1.4em]
    \item each $k_i(\cdot)$ is continuous for almost all $t \ge t_0$ (discontinuities have Lebesgue measure zero);
    \item there exists $\underline{\kappa} > 0$ such that $k_i(t) \in [\underline{\kappa}, \infty)$ for $i \in S$, and $k_j(t) \equiv 0$ for $j \notin S$, $\forall\, t \ge t_0$.
\end{enumerate}
\end{theorem}
	\begin{proof}
		We begin by defining the error vector $\mathbf e(t):= \mathbf x(t)-\mathbf x_d $ associated with the system dynamics \eqref {system_1} with the control input given in eqn. \eqref{u}. The corresponding error dynamics is given by
		\begin{equation}
			\dot{\mathbf e}(t)
			= -\bigl(L(t)+K(t)\bigr)\,\mathbf e(t).
			\label{error_2}
		\end{equation}
		We now show that \(\mathbf{e}(t)\to\mathbf{0}\) as \(t\to\infty\), and consequently \(\mathbf{x}(t)\to\mathbf{x}_d\).
Before proceeding with the proof, we introduce a few more notations and definitions used therein. Let $V_0 \subseteq S$ denote the subset of nodes in $S$ that have at least one immediate neighbor in the receiver set $R$ within the graph $G_\delta^\infty$. Note that  $G_\delta^\infty$ is the graph formed by the union of all the $\delta$-arcs over the  intervals \([sT,\,(s+1)T)\) for \(s=0,1,2,\ldots\). Let us define 
		\begin{align}
			h^e(t) &:= \max_{i \in R} |e_i(t)|, \label{h_e}\\
			c^e(t)&:=\max_{i \in S} |e_i(t)|
		\end{align}
From Theorem~\eqref{thm_1}, we have $k_i(t) > 0$ for $i \in S$ and $k_i(t) = 0$ for all $i \in V \setminus S=R$. Consequently, the absolute error dynamics share the same structure as the absolute state dynamics with zero control input in eqn.~\eqref{eq:abs-dyn}, except that $|x_i|$ is replaced by $|e_i(t)|$. For nodes in the root set $S$, an additional term $-k_i(t)|e_i(t)|$ appears due to the matrix $K(t)$ in eqn.~\eqref{error_2}. Hence, for any $i \in S$, the absolute error dynamics is given as
\begin{align}
    \frac{d}{dt}|e_i(t)| 
    = \sum_{j \in \mathcal{N}_i} |a_{ij}(t)|
      \big(\tilde{\theta}_{ij}|e_j(t)| - |e_i(t)|\big)
      - k_i(t)|e_i(t)|.
    \nonumber
\end{align}
where $\tilde{\theta}_{ij}:=\operatorname{sgn}(e_i(t))\,\operatorname{sgn}(a_{ij}(t))\,\operatorname{sgn}(e_j(t))\in \{-1,0,1\}$.
Then the upper Dini derivatives of $h^e(t)$ and $c^e(t)$ are obtained in the manner similar to that described in Lemma~\ref{max_error_dynamics}, and the resulting expressions are
\begin{align}
	D^+ h^e(t) &\leq M_0 N_s \big(c^e(t) - h^e(t)\big), \label{Dini_h_e}\\
	D^+ c^e(t) &\leq -\underline{\kappa}\, c^e(t). \label{Dini_c_e}
\end{align}	
Assuming $t_0 = 0$ for simplicity and applying Grönwall’s inequality in the same manner as in Lemma~\ref{k_times}, we obtain that, for almost all $t \in [sK_0,\,(s + 1)K_0]$, the following bounds hold:
        \begin{align}
    c^e(t) &\le e^{-\underline{\kappa}(t - sK_0)}\,c^e(sK_0)
        \;=\; \beta(t)\,c^e(sK_0), \label{c_11}\\[3pt]
    h^e(t) &\le h^e(sK_0) + c^e(sK_0). \label{c_12}
\end{align}
where \( s, k \in \mathbb{Z}_{\ge 0} \), \( \beta(t) :=e^{-\underline{\kappa}(t - sK_0)}\) and \( K_0 = kT \), represents a large time interval composed of $k$ no. of sub intervals of length \( T \).\\     
We now establish, in successive steps, upper bounds on the absolute error \( |e_i(t)| \) for all agents \( i = 1, 2, \dots, N \) over the interval \([sK_0, (s+1)K_0)\) with $K_0=d_0 T$, where \( d_0 \) denotes the longest path length from the root set $S$ to any other node in the in the graph \( G_\delta^\infty \). Note that we assume each subinterval of length $T$ allows the propagation of influence one step forward along the communication path. Hence, to ensure that the influence from $S$ reaches even the most distant node (at distance $d_0$), we consider a time interval long enough for information to traverse all $d_0$ steps, i.e., $K_0 = d_0 T$. \\
\textbf{Step 1:}~(Root node set bounds):    
Let \(I_m := [\,j_m T,\,(j_m+1)T\,) \subseteq [\,sK_0,\,(s+1)K_0\,)\), \(m=1,\dots,d_0\), denote the \(d_0\) disjoint subintervals of length \(T\) such that 
\([\,sK_0,\,(s+1)K_0\,) = \bigcup_{m=1}^{d_0} I_m\).  
From \eqref{c_11}, for all \(i \in V_0\),
\[
|e_i(t)| \leq c^e(t) \leq \beta(t) c^e(sK_0),
\quad \text{a.a. } t\in [\,sK_0,\,(s+1)K_0\,].
\]
\textbf{Step 2}~(Follower set bounds):  Let us define
\begin{align*}
			\mathcal{V}_1 :=& 
			\Big\{\, j \in R \;\Big|\; 
			\text{there exists a \(\delta\)-arc from a node in $V_0$ to} \notag\\
			&\text{\(j \in R\) on  time interval $[j_1T,(j_1+1)T)$} \Big\}.
		\end{align*}
Let \(i_1\in\mathcal{V}_1\), \(z_i:=|e_i(t)|\), and 
\(Y_{i_1}(t):=\sum_{j\in\mathcal{N}_{i_1}}|a_{i_1j}(t)|\).  Then, for a.a. \(t\in [j_1T,(j_1+1)T]\),
\begin{align}
\dot z_{i_1}(t)&\le 
\sum_{j\in\mathcal{N}_{i_1}\cap R}|a_{i_1j}(t)|(B_1-z_{i_1})\nonumber\\
&+\sum_{j\in\mathcal{N}_{i_1}\cap V_0}|a_{i_1j}(t)|(B_2-z_{i_1}), \label{step_0}
\end{align}
where \(B_1=h^e(sK_0)+c^e(sK_0)\), \(B_2=\beta(t) c^e(sK_0)\), and 
\(\widehat{Y}_{i_1}(t):=Y_{i_1}(t)-\sum_{j\in\mathcal{N}_{i_1}\cap V_0}|a_{i_1j}(t)|\).
Applying Grönwall’s inequality yields
\begin{equation}
z_{i_1}((j_1+1)T)\le 
z_{i_1}(j_1T)e^{-I}+B_1\hat{I}_Y+B_2I_a,
\label{z_1}
\end{equation}
where
\[
\begin{aligned}
I &= \!\int_{j_1T}^{(j_1+1)T}\! Y_{i_1}(t)\,dt,\nonumber\\ 
\hat{I}_Y &= \!\int_{j_1T}^{(j_1+1)T}\! \widehat{Y}_{i_1}(t)\,e^{-\!\int_t^{(j_1+1)T}\! Y_{i_1}(\tau)\,d\tau}\,dt, \\
I_Y &= \!\int_{j_1T}^{(j_1+1)T}\! Y_{i_1}(t)\,e^{-\!\int_t^{(j_1+1)T}\! Y_{i_1}(\tau)\,d\tau}\,dt, \\[3pt]
I_a &= \!\int_{j_1T}^{(j_1+1)T}\!\!\Bigg(\sum_{j\in\mathcal{N}_{i_1}\cap V_0}\!|a_{i_1j}(t)|\,e^{-\!\int_t^{(j_1+1)T}\! Y_{i_1}(\tau)\,d\tau}\Bigg)\!dt.
\end{aligned}
\]
Substituting the values of \(B_1\) and \(B_2\) back in eqn. \eqref{z_1} and using the fact that $z_{i_1}(j_1T) \leq h^e(sK_0)+c^e(sK_0)$ for a.a. $t \in [sK_0,\,(s + 1)K_0]$, $\beta(t) \leq 1$ and from direct computation $I_Y = \hat{I}_Y + I_a$, $I_Y = 1 - e^{-I},$ we obtain
		\begin{equation}
			z_{i_1}((j_1+1)T) \;\leq\; h(sK_0)\,(1 - I_a) + c(sK_0).\\
			\label{step_1}
		\end{equation} 
Moreover, by definition,
		\begin{align}
			I_a
			&= \int_{j_1T}^{(j_1+1)T}
			\Bigg(\sum_{j\in \mathcal{N}_{i_1}  \cap V_0} |a_{i_1 j}(t)|\Bigg)
			e^{-\int_t^{(j_1+1)T} Y_{i_1}(\tau)\,d\tau}\,dt \nonumber\\[1ex]
			&= \int_{j_1T}^{(j_1+1)T}
			\Bigg(\sum_{j\in \mathcal{N}_{i_1}  \cap V_0} |a_{i_1 j}(t)|\Bigg)\,
			e^{-\int_t^{(j_1+1)T} \widehat{Y}_{i_1}(\tau)\,d\tau} \nonumber\\
			&\qquad \times e^{-\int_t^{(j_1+1)T} 
				\sum_{j\in \mathcal{N}_{i_1}  \cap V_0}|a_{i_1 j}(\tau)|\,d\tau}\,dt \nonumber\\
			&\geq e^{-M_0 (N - N_s - 1) T}
			\Bigg(1 - e^{-\int_{j_1T}^{(j_1+1)T} \sum_{j\in \mathcal{N}_{i_1}  \cap V_0} |a_{i_1 j}(t)|\,dt}\Bigg) \nonumber\\[1ex]
			&\ge e^{-M_0 (N - N_s - 1) T}\,\big(1 - e^{-\delta T}\big) =: \zeta. \label{step_12}
		\end{align}
		where we have used the fact that 
		\begin{align*}
			&\Bigg(1 - e^{-\int_{j_1T}^{(j_1+1)T} \sum_{j\in V_0} |a_{i_1 j}(t)|\,dt}\Bigg) \geq (1 - e^{-\delta T}\big) \text{, and }\nonumber\\
            &e^{-\int_t^{(j_1+1)T} \widehat{Y}_{i_1}(\tau)\,d\tau} \geq e^{-M_0 (N - N_s - 1) T}
		\end{align*}
		where $N$ is the total number of agents in the graph $\mathcal{G}_{A(t)}$, $N_s$ is the number of agents in the root node set $S$, and $(N - N_s)$ is the number of receiver nodes in $R$. The term $(N - N_s - 1)$ denotes the maximum possible number of incoming $\delta$–arcs to a node $i_1 \in \mathcal{V}_1$ from its neighbors in $R$, excluding node $i_1$ itself (since  self-loops are 0). Now substituting eqn. \eqref{step_12} into  eqn. \eqref{step_1}, we get
		\begin{equation}
			z_{i_1}((j_1+1)T) \;\leq\; h^e(sK_0)\,(1 - \zeta) + c^e(sK_0).\\
			\label{bound_1}
		\end{equation}
		Now, using the inequality similar to eqn. \eqref{step_0}, we obtain that for a.a. $t \in [(j_1 + 1)T,\,(s + 1)K_0]$,
		\begin{align*}
			\dot{z}_{i_1}(t) &\leq  \sum_{j \in \mathcal{N}_{i_1}  \cap R} |a_{i_1 j}(t)|  z_j(t)+\sum_{j \in \mathcal{N}_{i_1}  \cap S} |a_{i_1 j}(t)|  z_j(t)\nonumber\\
			&-\sum_{j \in \mathcal{N}_{i_1}} |a_{i_1 j}(t)|z_{i_1}(t)\nonumber\\
			&\leq Y_{i_1}(t) (h^e(sK_0)+2c^e(sK_0))- Y_{i_1}(t)  z_{i_1}(t) 
            \end{align*}
            Applying Grönwall’s inequality, it turns out that for a.a. \(t \in \big[(j_1 + 1)T,\,(s + 1)K_0\big]\),
     \begin{equation}
			 z_{i_1}(t) \leq h^e(sK_0) \cdot \left(1 - \xi_0 \zeta \right)
			+ c^e(sK_0) (2-\xi_0),
			\label{leader_bound_2}
		\end{equation}
		where we have used the fact that $e^{ - \int_{(j_1 + 1)T}^{t} Y_{i_1}(\tau)\, d\tau } \geq e^{-(N-1) M_0 K_0} \, =: \xi_0 $.\\
	\textbf{Step 3:} Continuing the similar analysis, we define
		\begin{align*}
			\mathcal{V}_2 :=& 
			\Big\{\, j \in R \;\Big|\; 
			\text{there exists a \(\delta\)-arc from $\mathcal{V}_1$ to} \notag\\
			&\text{\(j \in R\) on  time interval $[j_2T,(j_2+1)T)$} \Big\}.
		\end{align*}
		Let $i_2 \in \mathcal{V}_2$, and  repeating with the previous analysis as given in step 2, we get that for a.a. $t \in [(j_2+1)T,(s+1)K_0]$, 
		\begin{equation}
			z_{i_2}(t) \leq h^e(sK_0)\big(1-\xi_0^2 \zeta \mathcal{X}\big) 
			+ c^e(sK_0)\big(2 - \xi_0^2 \mathcal{X})\big), 
		\end{equation}
Recall that $d_0$ is the longest path length of $G_\delta^\infty$. We can proceed with the similar analysis on time intervals \( [j_m T, (j_m + 1) T) \) for \( m=3, \ldots, d_0 \),  
and define subsets \( \mathcal{V}_3, \ldots, \mathcal{V}_z \) for some \( z \leq d_0 \), respectively,  
until we obtain  $V \;=\; \Big( \bigcup_{i=1}^{z} \mathcal{V}_i \Big) \cup V_0$.
 Thus, for a.a. $t\in[(j_{d_0}+1)T,(s+1)K_0]$, we get
\begin{align}
z_{i_{d_0}}(t)
&\le h^e(sK_0)\left(1-\xi_0^{d_0}\zeta\mathcal{X}^{d_0-1}\right)
+\nonumber\\
&c^e(sK_0)\left(2-\xi_0^{d_0}\mathcal{X}^{d_0-1}\right)
\end{align}
Since  $0<\xi_0\,\zeta\,\mathcal X\le 1$, we have $\xi_0^{d_0}\zeta\mathcal X^{d_0-1}\le \xi_0^{d_0-1}\zeta\mathcal X^{d_0-2}\le \cdots \le \xi_0\zeta$. Hence the upper bound increases with each iteration and we can write 
		\begin{align}
			h^e((s+1)K_0) 
			&\leq h^e(sK_0)\Bigl(1 - \xi_0^{d_0}\,\zeta\,\mathcal{X}^{d_0-1}\Bigr) \notag \\
			&\quad + c^e(sK_0)\Bigl(2 - \xi_0^{d_0}\,\mathcal{X}^{d_0-1}\Bigr).
			\label{ar}
		\end{align}
		Since $s$ is arbitrary chosen in eqn. \eqref{ar}, we have 
		\begin{align}
			h^e(nK_0)
			&\le h^e\bigl((n-1)K_0\bigr)\Bigl(1-\xi_0^{d_0}\zeta\mathcal X^{\,d_0-1}\Bigr)
			+\nonumber\\ &c^e\bigl((n-1)K_0\bigr)\Bigl(2-\xi_0^{d_0}\mathcal X^{\,d_0-1}\Bigr).
            \label{he}
		\end{align}
     for any $n\in\mathbb{Z}_{\ge1}$.
		Let $\alpha_1:=1-\xi_0^{d_0}\zeta\mathcal X^{\,d_0-1}$ and $\alpha_2:=\xi_0^{d_0}\mathcal X^{\,d_0-1}$,
		and write $h^e_n:=h^e(nK_0)$, $c^e_n:=c^e(nK_0)$ for notational simplicity. Then from eqn. \eqref{he}, we get
		\begin{equation}
		h^e_n \le \alpha_1 h^e_{n-1} + (2-\alpha_2)c^e_{n-1},\qquad n\in\mathbb{Z}_{\ge1}.
        \label{he1}
		\end{equation}
		Let  $c^e_n\downarrow c^e_\infty$ and $h^e_n\downarrow h^e_\infty$ as $n\to\infty$. Hence, we get 
		\begin{equation}
		h^e_\infty \le \frac{2-\alpha_2}{\,1-\alpha_1\,}c^e_\infty.
        \label{infty}
		\end{equation}
		Moreover, from eqn. \eqref{c_12}, for a.a. $t\in[nK_0,(n+1)K_0]$ we have $h^e(t)\le h^e_n + c^e_n$, hence
		\begin{align}
		\lim_{t\to\infty} h^e(t)
		&\le \lim_{n\to\infty}h^e_n + c^e_n=h^e_\infty+c^e_\infty\nonumber\\
		&\le \Bigl(\frac{2-\alpha_2}{1-\alpha_1}+1\Bigr)c^e_\infty \quad \text{from eqn. \eqref{infty}}\nonumber\\
		&= \frac{3-\alpha_2-\alpha_1}{1-\alpha_1}\;c^e_\infty.
        \label{t}
		\end{align}
We also know from eqn. \eqref{c_11} that for any $s\in\mathbb{Z}_{\ge0}$ and a.a. $t \in [sK_0,(s+1)K_0]$,
\[
c^e(t)\ \le\ c^e(sK_0)\,e^{-\underline\kappa\,(t-sK_0)}.
\]
In particular, for $t=(s+1)K_0$, we obtain 
\[
c^e\bigl((s+1)K_0\bigr)\ \le\ e^{-\underline\kappa K_0}\,c^e(sK_0)
=\tilde{\beta}\,c^e(sK_0),
\]
where  $ \tilde{\beta}:= e^{-\underline\kappa K_0}\in(0,1)$. Since $s$ is arbitrary, we can write  $c^e_n\le \tilde{\beta} c^e_{n-1}$ for $n \in \mathbb{Z}_{\geq1}$. This implies $c^e_n\le (\tilde{\beta})^{\,n}c^e_0$ and $c^e_n \to0$ as $n\to\infty$ i.e, $c^e_\infty=0$. Hence, from eqn. \eqref{t}, we conclude that $\lim_{t\to\infty} h^e(t)=0$. This implies that the absolute value of error converges to $0$ and hence $\mathbf{x}(t)$ converges to $\mathbf{x}_d$ as $t\to\infty$ . 
		\end{proof}

\section{Numerical Simulations} \label{sim}
\begin{figure}[t]
  \centering
  \includegraphics[width=0.8\columnwidth, keepaspectratio]{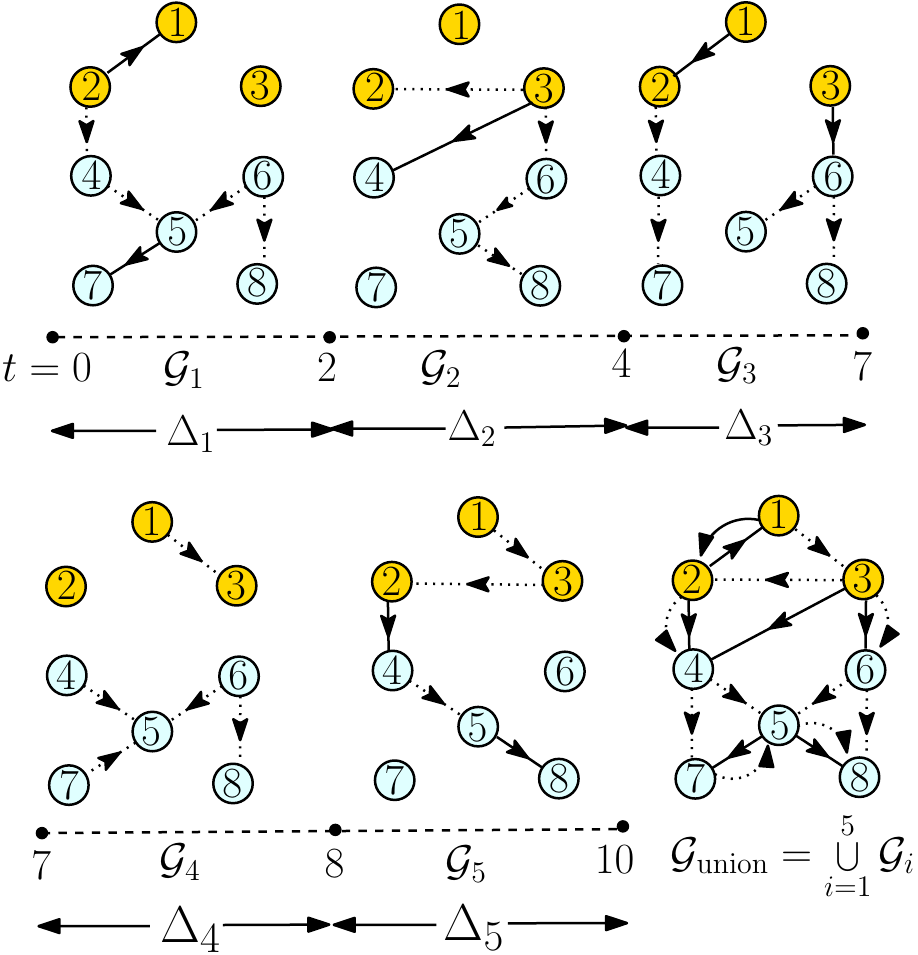}\vspace{-0.3 cm}
  \caption{Sequence of graphs $\{\mathcal{G}_1, \mathcal{G}_2, \ldots, \mathcal{G}_5 \}$  over the interval $[0, 10)$.}
  \label{fig:single}
\end{figure}

We consider a time-varying signed network of $8$ agents satisfying the conditions of Theorem~\ref{thm_1}. 
The switching topology is generated by a cyclic activation of five distinct graphs 
$\{\mathcal{G}_1, \ldots, \mathcal{G}_5\}$, each active for a dwell time $\Delta_i$ (in seconds). 
The set of dwell times is given by 
$(\Delta_1,\Delta_2,\Delta_3,\Delta_4,\Delta_5) = (2,2,3,1,2)\,$ for each $\mathcal{G}_i$. The graphs switch in a rotating cyclic pattern, i.e., 
$\mathcal{G}_1 \!\rightarrow\! \mathcal{G}_2 \!\rightarrow\! \mathcal{G}_3 \!\rightarrow\! \mathcal{G}_4 \!\rightarrow\! \mathcal{G}_5$, 
followed by 
$\mathcal{G}_2 \!\rightarrow\! \mathcal{G}_3 \!\rightarrow\! \mathcal{G}_4 \!\rightarrow\! \mathcal{G}_5 \!\rightarrow\! \mathcal{G}_1$, 
then 
$\mathcal{G}_3 \!\rightarrow\! \mathcal{G}_4 \!\rightarrow\! \mathcal{G}_5 \!\rightarrow\! \mathcal{G}_1 \!\rightarrow\! \mathcal{G}_2$, 
and so on, repeating periodically over time. Figure~\ref{fig:single} illustrates the sequence of activated graphs over the interval $[0,10)$, along with the corresponding union graph $\mathcal{G}_{\mathrm{union}}$, which represents the union of all $\delta$-arcs appearing within this period. We have assumed that each edge in $\mathcal{G}_i$ $(i=1,\ldots,5)$ is a $\delta$-arc.

\begin{figure}[h]
		\centering
		\includegraphics[width=0.4\textwidth, height=0.4\textheight,keepaspectratio]{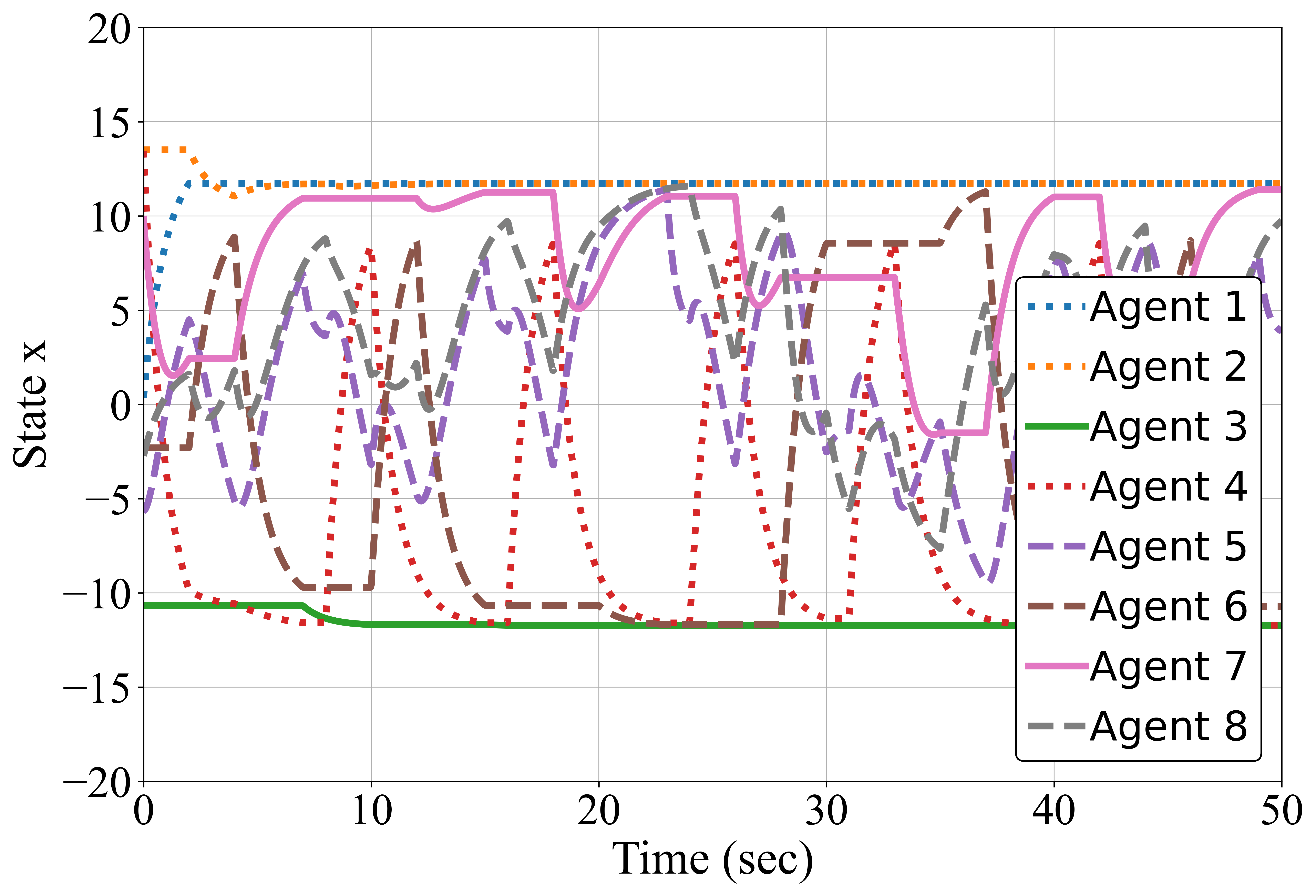}\vspace{-0.4 cm}
		\caption{Opinion evolution for system dynamics \eqref{system_1} with $\mathbf u(t)=0$.}
		\label{fig:figure6}   
	\end{figure}

This switching pattern yields a uniformly quasi–strongly $\delta$-connected graph over every time interval $[t, t+T)$ of length $T = 10~sec$. The network also contains a fixed node set $S = {1, 2, 3}$ that acts as the root node set in the union graph over each such interval and remains persistently structurally balanced. Figure~\ref{fig:figure6} shows the evolution of agent states without control input (i.e., $\mathbf{u}(t) \equiv 0$), where the root agents in $S$ exhibit polarization, and the non-root agents’ opinions oscillate between the polarized values of the root nodes.

\begin{figure}[h]
		\centering
\includegraphics[width=0.39\textwidth, height=0.39\textheight,keepaspectratio]{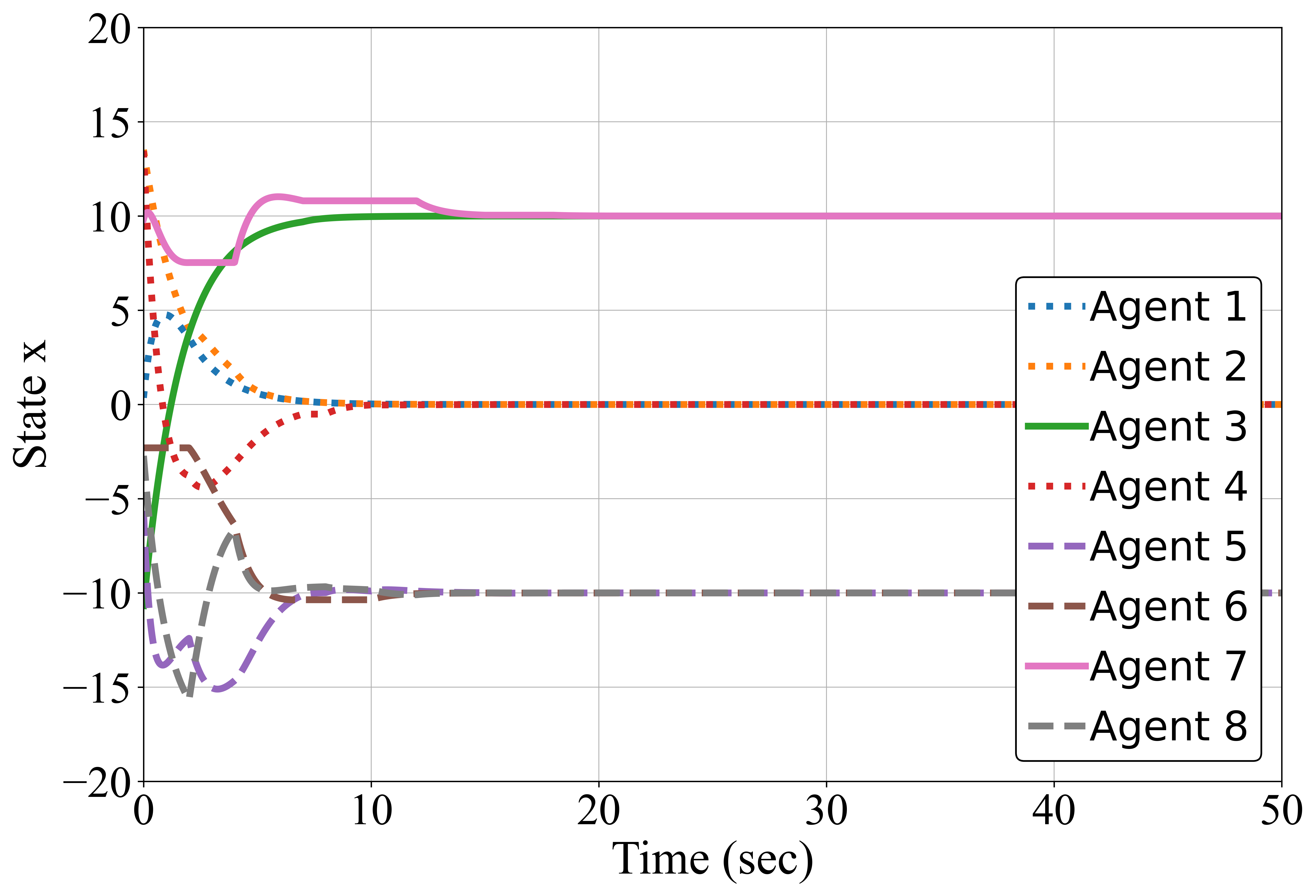}\vspace{-0.4 cm}
		\caption{Desired opinion clustering of agents with  control input $\mathbf{u}(t)$.}\vspace{-0.5 cm}
		\label{fig:figure5}   
	\end{figure}

We now set the target opinion vector as $\mathbf{x}_d = [0, 0, 10, 0, -10, -10, 10, -10]^\top$ and design the control input $\mathbf{u}(t)$ as described in Theorem~\ref{thm_1}. The controller parameter $k_i(\dot)$ is chosen as $k_i(t) = 0.6$ for $i \in S$ and $k_i(t) = 0$ for $i \in V \setminus S$, $\forall t \geq 0$. This choice satisfies conditions (P1)–(P3) of Theorem~\ref{thm_1}, ensuring a valid control law. Under this control, all agent states converge to the desired target state $\mathbf{x}_d$, i.e., $\mathbf{x}(t) \to \mathbf{x}_d$ as shown in Fig.~\ref{fig:figure5}.

\vspace{-0.3 cm}

\section{Discussion}\label{disc}
     % \vspace{-0.1 cm}
Public opinion in modern information ecosystems is increasingly shaped by exogenous factors such as algorithmic curation, biased media coverage, and coordinated misinformation campaigns. High-impact events (e.g., the 2016 U.S. presidential election and the COVID-19 pandemic) demonstrate how influential agents (politicians, journalists, celebrities) can exploit social media to spread misinformation and amplify polarization. These phenomena motivate the need for principled intervention strategies that can steer collective opinion toward informed and socially beneficial outcomes. In this work, we adopt a control-theoretic perspective in which such interventions are modeled as external control inputs, representing targeted messaging by trusted institutions (e.g., public health agencies) that drive the network state toward a desired opinion profile, as illustrated in Fig.~\ref{fig:figure5}.
\vspace{-0.2 cm}
	\section{Conclusion}\label{con}
In this work, we have developed a control-theoretic framework for targeted opinion formation in networks with time-varying and signed interactions. Building on existing models, we designed a decentralized controller that leverages knowledge of the network topology to steer every agent’s opinion toward a prescribed target state. Our main theoretical result stated in Theorem \ref{thm_1} shows that if the network is uniformly quasi-strongly $\delta$-connected and contains a fixed set of nodes called the root node, which is persistently structurally balanced, then applying the proposed control law guarantees exponential convergence of every agent’s state to its target state. This convergence is rigorously proven via a novel analysis using the upper Dini derivative, Grönwall-type inequalities, and recursive error propagation bounds. These findings demonstrate that external interventions (public information campaigns, fact-checking systems, or algorithmic adjustments) can reliably shape collective beliefs in evolving social networks. Future work will explore extensions to networks in the presence of delays and noise.

	\bibliographystyle{IEEEtran}
	\bibliography{refs}

\end{document}